\documentclass[conference]{IEEEtran}
\usepackage{amsmath}
\usepackage{amssymb}
\usepackage{amsthm}
\usepackage{enumerate}
\usepackage[T1]{fontenc}
\usepackage{hyperref}
\usepackage{mleftright}

\newcommand{\E}[1]{\mathrm{E}[#1]}
\newcommand{\binaryemptyleft}{\mkern-\medmuskip{}}
\newcommand{\clmref}[1]{Claim~\ref{#1}}
\newcommand{\propref}[1]{Proposition~\ref{#1}}
\newcommand{\reals}{\mathbb{R}}
\newcommand{\secref}[1]{Section~\ref{#1}}
\newcommand{\set}[1]{\mathcal{#1}}
\newcommand{\thmref}[1]{Theorem~\ref{#1}}
\newcommand{\vthinspace}{\hspace{0.083em}}
\newtheorem{Proposition}{Proposition}
\newtheorem{Theorem}[Proposition]{Theorem}

\begin{document}
\title{Gambling and R\'enyi Divergence}

\author{\IEEEauthorblockN{C\'edric Bleuler, Amos Lapidoth, and Christoph Pfister}
\IEEEauthorblockA{Signal and Information Processing Laboratory\\
ETH Zurich, 8092 Zurich, Switzerland\\
Email: cbleuler@student.ethz.ch; \{lapidoth,pfister\}@isi.ee.ethz.ch}}

\maketitle

\begin{abstract}
For gambling on horses, a one-parameter family of utility functions is proposed, which contains Kelly's logarithmic criterion and the expected-return criterion as special cases.
The strategies that maximize the utility function are derived, and the connection to the R\'enyi divergence is shown.
Optimal strategies are also derived when the gambler has some side information; this setting leads to a novel conditional R\'enyi divergence.
\end{abstract}

\section{Introduction}

Consider a horse race with $m \ge 1$ horses $1,\ldots,m$, where the $i$-th horse wins with probability $p_i > 0$, and on which a bookie offers odds $o_i > 0$ for $1$.
A gambler spends all her wealth $\gamma_0 > 0$ to place bets on the horses.
Let $b_i \ge 0$ denote the fraction of $\gamma_0$ that the gambler bets on the \mbox{$i$-th} horse.
Let the random variable $X$ denote the winning horse, and define the wealth relative $S$ as
\begin{IEEEeqnarray}{l}
S \triangleq b_X \vthinspace o_X,\IEEEeqnarraynumspace
\end{IEEEeqnarray}
so the gambler's wealth after one race is $\gamma_1 = \gamma_0 \vthinspace S$.

Kelly \cite{Kelly} observed that in the setting where the odds and winning probabilities remain constant over many independent races and the gambler keeps investing all her wealth with the same relative allocation $b_1,\ldots,b_m$, the exponential rate of growth of the gambler's wealth tends to $\E{\log S}$ with probability one, i.e.,
\begin{IEEEeqnarray}{l}
\lim_{n \to \infty} \frac{1}{n} \log \frac{\gamma_n}{\gamma_0} = \E{\log S},\IEEEeqnarraynumspace\label{eq:exprateofgrowth}
\end{IEEEeqnarray}
where $\gamma_n$ denotes the gambler's wealth after $n$ horse races, and
$\log \vthinspace (\cdot)$ denotes the base-2 logarithm.
The RHS of \eqref{eq:exprateofgrowth} is known as the doubling rate \cite[Section~6.1]{CoverThomas}.

In this paper, we seek betting strategies that maximize
\begin{IEEEeqnarray}{l}
U_\beta \triangleq \frac{1}{\beta} \log \E{S^\beta},\IEEEeqnarraynumspace
\end{IEEEeqnarray}
where $\beta \in \reals \setminus \{0\}$ is a parameter.
This family of utility functions generalizes several important cases:

\begin{enumerate}[a)]
\item In the limit as $\beta$ tends to zero, $U_\beta$ tends to the doubling rate $\E{\log S}$, and we recover Kelly's result: irrespective of the odds, the optimal strategy is proportional betting, i.e., choosing $b_i = p_i$ for $i \in \{1,\ldots,m\}$; see \propref{prop:allmoneyzero}.
\item If $\beta = 1$, then maximizing $U_\beta$ is equivalent to maximizing $\E{S}$, the expected return, and it is optimal to put all the money on a horse that maximizes $p_i \vthinspace o_i$; see \propref{prop:allmoneydegen}.
\item In general, if $\beta \ge 1$, then it is optimal to put all the money on one horse; see \propref{prop:allmoneydegen}.
This is risky: if that horse loses, the gambler will be broke.
\item In the limit as $\beta$ tends to $+\infty$, it is optimal to put all the money on a horse that maximizes $o_i$, ignoring the winning probabilities.
This strategy maximizes the best-case payoff; see \propref{prop:allmoneyposinf}.
\item In the limit as $\beta$ tends to $-\infty$, it is optimal to choose $b_i = c / o_i$ for $i \in \{1,\ldots,m\}$, where $c$ is the normalizing constant defined in \eqref{eq:defc} ahead.
This strategy maximizes the worst-case payoff and is completely risk-free: irrespective of which horse wins, $S = c$; see \propref{prop:allmoneyneginf}.
\end{enumerate}

Our utility function has the following underlying structure: it is the logarithm of a (weighted) power mean \cite{CSMC,MeansAndInequalities}:
\begin{IEEEeqnarray}{l}
U_\beta = \log \mleft[\vthinspace\sum_{i = 1}^m p_i \vthinspace (b_i \vthinspace o_i)^\beta\mright]^\frac{1}{\beta}.\IEEEeqnarraynumspace\label{eq:ubetapowermean}
\end{IEEEeqnarray}
For $\beta \in \{-\infty, 0, 1, \infty\}$, the power mean is equal to the minimum, the geometric mean, the arithmetic mean, and the maximum of the set $\{b_i \vthinspace o_i\}_{i = 1}^m$, respectively.
Campbell \cite{CampbellA,CampbellB} used a cost function with a structure similar to \eqref{eq:ubetapowermean} to provide an operational meaning to the R\'enyi entropy in source coding.
Other information-theoretic examples of exponential moments were studied in \cite{ExpMoments}.
The utility function $U_\beta$ can be motivated by risk aversion models in finance theory \cite[(8)]{Soklakov}.

Our main result is \thmref{thm:allmoneyregular}, which shows that for $\beta < 1$, $U_\beta$ can be written as the sum of three terms; the central role is played by the R\'enyi divergence.
After dealing with the other values of $\beta$, we treat in \thmref{thm:sideinfo} the situation where the gambler, prior to placing her bets, observes some side information.
This analysis features a novel conditional R\'enyi divergence, whose properties are studied in Propositions~\ref{prop:conddivprop} and~\ref{prop:uncondleconddiv}.
In \propref{prop:fairsuperfair} and \thmref{thm:partmoneyregular}, we study the situation where the gambler invests only part of her money.

The rest of this paper is structured as follows:
In \secref{sec:preliminaries}, we recall the R\'enyi divergence and define a conditional R\'enyi divergence, and in \secref{sec:results}, we present our results; all proofs are deferred to \secref{sec:proofs}.

\section{Preliminaries}
\label{sec:preliminaries}

The following definitions are for probability mass functions (PMFs); the definitions for probability vectors are analogous.
When clear from the context, we often omit sets and subscripts: for example, we write $\sum_x$ for $\sum_{x \in \set{X}}$ and $p(x)$ for $p_X(x)$.
The R\'enyi divergence of order $\alpha$ between two PMFs $p_X$ and $q_X$ \cite{RenyiDivergence} is defined for positive $\alpha$ other than one as
\begin{IEEEeqnarray}{l}
D_\alpha(p_X\|q_X) \triangleq \frac{1}{\alpha - 1} \log \sum_x p(x)^\alpha \vthinspace q(x)^{1 - \alpha}.\IEEEeqnarraynumspace
\end{IEEEeqnarray}
Its properties are studied in \cite{VanErvenHarremoes}.

Let $p_Y$ be a PMF, and let $p_{X|Y}$ and $q_{X|Y}$ be conditional PMFs.
We define the conditional R\'enyi divergence of order $\alpha$ for positive $\alpha$ other than one as
\begin{IEEEeqnarray}{rCl}
\IEEEeqnarraymulticol{3}{l}{D_\alpha(p_{X|Y}\|q_{X|Y}|p_Y)}\IEEEeqnarraynumspace\nonumber\\[-0.5ex]\quad
&\triangleq& \frac{\alpha}{\alpha - 1} \log \sum_y p(y) \mleft[\sum_x p(x|y)^\alpha \vthinspace q(x|y)^{1 - \alpha}\mright]^\frac{1}{\alpha}.\IEEEeqnarraynumspace
\end{IEEEeqnarray}
This definition differs from other definitions of the conditional R\'enyi divergence \cite[(6) and (8)]{VerduAlphaMutual}.
Some of its properties are presented in Propositions~\ref{prop:conddivprop} and \ref{prop:uncondleconddiv} ahead.

\section{Results}
\label{sec:results}

We first analyze the situation where the gambler invests all her money, i.e., where $b \triangleq (b_1,\ldots,b_m)$ is a probability vector.
(A probability vector is a vector with nonnegative components that add up to one.)
As in \cite[Section~10.3]{Moser}, define
\begin{IEEEeqnarray}{l}
c \triangleq \mleft[\vthinspace\sum_{i = 1}^m \frac{1}{o_i}\mright]^{-1},\IEEEeqnarraynumspace\label{eq:defc}
\end{IEEEeqnarray}
the probability vector $p \triangleq (p_1,\ldots,p_m)$, and the probability vector $r \triangleq (r_1,\ldots,r_m)$, where for $i \in \{1,\ldots,m\}$,
\begin{IEEEeqnarray}{l}
r_i \triangleq \frac{c}{o_i}.\IEEEeqnarraynumspace\label{eq:defr}
\end{IEEEeqnarray}

\begin{Theorem}
\label{thm:allmoneyregular}
Let $\beta \in (-\infty,0) \cup (0,1)$, and let $b$ be a probability vector.
Then,
\begin{IEEEeqnarray}{l}
\frac{1}{\beta} \log \E{S^\beta} = \log c + D_{\!\frac{1}{1 - \beta}}(p\|r) - D_{1 - \beta}(g\|b),\IEEEeqnarraynumspace\label{eq:thmallregsplit}
\end{IEEEeqnarray}
where for $i \in \{1,\ldots,m\}$,
\begin{IEEEeqnarray}{l}
g_i \triangleq \frac{p_i^{\frac{1}{1 - \beta}} o_i^{\frac{\beta}{1 - \beta}}}{\sum_{j = 1}^m p_j^{\frac{1}{1 - \beta}} o_j^{\frac{\beta}{1 - \beta}}}.\IEEEeqnarraynumspace\label{eq:thmallregdefgi}
\end{IEEEeqnarray}
Thus, the choice $b = g$ uniquely maximizes $\frac{1}{\beta} \log \E{S^\beta}$ among all probability vectors $b$.
\end{Theorem}

We see from \thmref{thm:allmoneyregular} that if $\beta \in (-\infty,0) \cup (0,1)$, then our utility function can be written as the sum of three terms:
\begin{enumerate}
\item The first term, $\log c$, depends only on the odds and is related to the fairness of the odds.
The odds are called subfair if $c < 1$, fair if $c = 1$, and superfair if $c > 1$.
\item The second term, $D_{\!\frac{1}{1 - \beta}}(p\|r)$, is related to the bookie's estimate of the winning probabilities.
It is zero if and only if the odds are inversely proportional to the winning probabilities.
\item The third term, $-D_{1 - \beta}(g\|b)$, is related to the gambler's estimate of the winning probabilities.
It is zero if and only if $b$ is equal to $g$.
\end{enumerate}

\begin{Proposition}
\label{prop:allmoneyzero}
Let $b$ be a probability vector.
Then,
\begin{IEEEeqnarray}{rCl}
\lim_{\beta \to 0} \frac{1}{\beta} \log \E{S^\beta} &=& \E{\log S}\IEEEeqnarraynumspace\label{eq:propallzeroa}\\[-1ex]
&=& \log c + D(p\|r) - D(p\|b).\IEEEeqnarraynumspace\label{eq:propallzerob}
\end{IEEEeqnarray}
\end{Proposition}

We see from \propref{prop:allmoneyzero} that in the limit as $\beta$ tends to zero, the doubling rate $\E{\log S}$ is recovered from our utility function.
Here, the analog of \eqref{eq:thmallregsplit} is \eqref{eq:propallzerob}; note that \eqref{eq:propallzerob} implies that $\E{\log S}$ is maximized if and only if $b$ is equal to $p$.

\begin{Proposition}
\label{prop:allmoneydegen}
Let $\beta \ge 1$, and let $b$ be a probability vector.
Then,
\begin{IEEEeqnarray}{l}
\frac{1}{\beta} \log \E{S^\beta} \le \log \max_{i \in \{1,\ldots,m\}} \bigl(\vthinspace p_i^{1 / \beta} o_i^{\vphantom{1 / \beta}}\bigr).\IEEEeqnarraynumspace\label{eq:propalldegen}
\end{IEEEeqnarray}
Equality in \eqref{eq:propalldegen} can be achieved by choosing
\begin{IEEEeqnarray}{l}
b_i = \begin{cases} 1 & \text{if $i = i^*$,}\\
0 & \text{otherwise,}\end{cases}\IEEEeqnarraynumspace\label{eq:propalldegenoptb}
\end{IEEEeqnarray}
where $i^* \in \{1,\ldots,m\}$ is such that
\begin{IEEEeqnarray}{l}
p_{i^*}^{1 / \beta} o_{i^*}^{\vphantom{1 / \beta}} = \max_{i \in \{1,\ldots,m\}} \bigl(\vthinspace p_i^{1 / \beta} o_i^{\vphantom{1 / \beta}}\bigr).
\end{IEEEeqnarray}
\end{Proposition}

We see from \propref{prop:allmoneydegen} that if $\beta \ge 1$, then it is optimal to bet on a single horse.
Unless $m = 1$, this is not the case when $\beta < 1$:
When $\beta < 1$, an optimal betting strategy requires placing a bet on every horse.
This follows from \thmref{thm:allmoneyregular} and our assumption that $p_i$ and $o_i$ are all positive.

\begin{Proposition}
\label{prop:allmoneyposinf}
Let $b$ be a probability vector.
Then,
\begin{IEEEeqnarray}{rCl}
\lim_{\beta \to +\infty} \frac{1}{\beta} \log \E{S^\beta} &=& \log \max_{i \in \{1,\ldots,m\}} b_i \vthinspace o_i\IEEEeqnarraynumspace\label{eq:propallposinfa}\\
&\le& \log \max_{i \in \{1,\ldots,m\}} o_i.\IEEEeqnarraynumspace\label{eq:propallposinfb}
\end{IEEEeqnarray}
Equality in \eqref{eq:propallposinfb} can be achieved by choosing
\begin{IEEEeqnarray}{l}
b_i = \begin{cases} 1 & \text{if $i = i^*$,}\\
0 & \text{otherwise,}\end{cases}\IEEEeqnarraynumspace\label{eq:propallposinfoptb}
\end{IEEEeqnarray}
where $i^* \in \{1,\ldots,m\}$ is such that $o_{i^*} = \max_{i \in \{1,\ldots,m\}} o_i$.
\end{Proposition}

\begin{Proposition}
\label{prop:allmoneyneginf}
Let $b$ be a probability vector.
Then,
\begin{IEEEeqnarray}{rCl}
\lim_{\beta \to -\infty} \frac{1}{\beta} \log \E{S^\beta} &=& \log \min_{i \in \{1,\ldots,m\}} b_i \vthinspace o_i\IEEEeqnarraynumspace\label{eq:propallneginfa}\\
&\le& \log c.\IEEEeqnarraynumspace\label{eq:propallneginfb}
\end{IEEEeqnarray}
Equality in \eqref{eq:propallneginfb} is achieved if and only if $b_i = c / o_i$ for all $i \in \{1,\ldots,m\}$.
\end{Proposition}

Our next result concerns the situation where the gambler observes some side information $Y$ before placing her bets.
To that end, we adapt our notation as follows:
Let $p_{XY}$ be the joint PMF of $X$ and $Y$.
(Recall that $X$ denotes the winning horse.)
Denote the range of $X$ and $Y$ by $\set{X}$ and $\set{Y}$, respectively.
We assume that $p(y) > 0$ for all $y \in \set{Y}$.
(Here, we do not assume that the winning probabilities $p(x)$ are positive.)
We view the odds as a function $o\colon \set{X} \to \reals_{> 0}$.
Define
\begin{IEEEeqnarray}{l}
c \triangleq \mleft[\vthinspace\sum_x \frac{1}{o(x)}\mright]^{-1},\IEEEeqnarraynumspace
\end{IEEEeqnarray}
and the PMF $r_X$ for $x \in \set{X}$ as
\begin{IEEEeqnarray}{l}
r_X(x) \triangleq \frac{c}{o(x)}.\IEEEeqnarraynumspace\label{eq:defrx}
\end{IEEEeqnarray}
(These definitions are equivalent to \eqref{eq:defc} and \eqref{eq:defr}, respectively.)
We continue to assume that the gambler invests all her wealth, so a betting strategy is now a conditional PMF $b_{X|Y}$.
The wealth relative $\tilde{S}$ is defined as
\begin{IEEEeqnarray}{l}
\tilde{S} \triangleq b_{X|Y}(X|Y) \vthinspace o(X).\IEEEeqnarraynumspace
\end{IEEEeqnarray}
The following theorem parallels \thmref{thm:allmoneyregular}:

\begin{Theorem}
\label{thm:sideinfo}
Let $\beta \in (-\infty,0) \cup (0,1)$.
Then,
\begin{IEEEeqnarray}{rCl}
\frac{1}{\beta} \log \E{\tilde{S}^\beta} &=& \log c + D_{\!\frac{1}{1 - \beta}}(p_{X|Y}\|r_X|p_Y)\IEEEeqnarraynumspace\nonumber\\
&& \binaryemptyleft - D_{1 - \beta}(g_{X|Y} \vthinspace g_Y\|b_{X|Y} \vthinspace g_Y),\IEEEeqnarraynumspace\label{eq:thmsidesplit}
\end{IEEEeqnarray}
where for $x \in \set{X}$ and $y \in \set{Y}$,
\begin{IEEEeqnarray}{rCl}
g(x|y) &\triangleq& \frac{p(x|y)^\frac{1}{1 - \beta} o(x)^\frac{\beta}{1 - \beta}}{\sum_{x'} p(x'|y)^\frac{1}{1 - \beta} o(x')^\frac{\beta}{1 - \beta}},\IEEEeqnarraynumspace\label{eq:thmsidedefgxy}\\
g(y) &\triangleq& \frac{p(y) \Bigl[\sum_{x'} p(x'|y)^\frac{1}{1 - \beta} o(x')^\frac{\beta}{1 - \beta}\Bigr]^{1 - \beta}}{\sum_{y'} p(y') \Bigl[\sum_{x'} p(x'|y')^\frac{1}{1 - \beta} o(x')^\frac{\beta}{1 - \beta}\Bigr]^{1 - \beta}}.\IEEEeqnarraynumspace\label{eq:thmsidedefy}
\end{IEEEeqnarray}
Thus, choosing $b_{X|Y} = g_{X|Y}$ uniquely maximizes $\frac{1}{\beta} \log \E{\tilde{S}^\beta}$ among all conditional PMFs $b_{X|Y}$.
\end{Theorem}

The conditional R\'enyi divergence $D_\alpha({\cdot\|\cdot}|\cdot)$ appearing in \thmref{thm:sideinfo} was defined in \secref{sec:preliminaries} and seems to be novel.
It is easy to see that $D_\alpha(p_X\|q_X|p_Y) = D_\alpha(p_X\|q_X)$ if $p_X$, $q_X$, and $p_Y$ are PMFs.
We now present some more properties:

\begin{Proposition}
\label{prop:conddivprop}
Let $\alpha \in (0,1) \cup (1,\infty)$, let $p_Y$ be a PMF, and let $p_{X|Y}$ and $q_{X|Y}$ be conditional PMFs.
Then,
\begin{IEEEeqnarray}{rCl}
0 &\le & D_\alpha(p_{X|Y}\|q_{X|Y}|p_Y)\IEEEeqnarraynumspace\label{eq:conddivnonzero}\\
&\le& D_\alpha(p_{X|Y} \vthinspace p_Y\|q_{X|Y} \vthinspace p_Y).\IEEEeqnarraynumspace\label{eq:conddivlefullcond}
\end{IEEEeqnarray}
\end{Proposition}

Because everything that can be achieved without side information can also be achieved with side information, comparing \thmref{thm:allmoneyregular} and \thmref{thm:sideinfo} suggests that $D_\alpha(p_X\|r_X) \le D_\alpha(p_{X|Y}\|r_X|p_Y)$, which is indeed the case:

\begin{Proposition}
\label{prop:uncondleconddiv}
Let $\alpha \in (0,1) \cup (1,\infty)$, let $p_{XY}$ be a joint PMF, and let $r_X$ be a PMF.
Then,
\begin{IEEEeqnarray}{l}
D_\alpha(p_X\|r_X) \le D_\alpha(p_{X|Y}\|r_X|p_Y).\IEEEeqnarraynumspace\label{eq:uncondleconddiva}
\end{IEEEeqnarray}
\end{Proposition}

Our last results treat the possibility that the gambler does not invest all her wealth.
(We only treat the setting without side information.)
Denote by $b_0$ the fraction of her wealth that the gambler does not use for betting.
Then, $b \triangleq (b_0,b_1,\ldots,b_m)$ is a probability vector, and the wealth relative $S_0$ is given by
\begin{IEEEeqnarray}{l}
S_0 \triangleq b_0 + b_X \vthinspace o_X.\IEEEeqnarraynumspace
\end{IEEEeqnarray}
If $c \ge 1$, then it is optimal to invest all the money:

\begin{Proposition}
\label{prop:fairsuperfair}
Assume $c \ge 1$, let $\beta \in \reals \setminus \{0\}$, and let $b$ be a probability vector with wealth relative $S_0$.
Then, there exists a probability vector $b'$ with wealth relative $S_0'$ satisfying $b_0' = 0$ and
\begin{IEEEeqnarray}{l}
\frac{1}{\beta} \log \E{S_0'^\beta} \ge \frac{1}{\beta} \log \E{S_0^\beta}.\IEEEeqnarraynumspace\label{eq:fairsuperfairbprime}
\end{IEEEeqnarray}
\end{Proposition}

On the other hand, if the odds are subfair, i.e., if $c < 1$, then investing all the money is not optimal in the case $\beta < 1$, as \clmref{clm:partmoneyregbzero} of the following theorem shows:

\begin{Theorem}
\label{thm:partmoneyregular}
Assume $c < 1$, let $\beta \in (-\infty,0) \cup (0,1)$, and let $b^*$ be a probability vector that maximizes $\frac{1}{\beta} \log \E{S_0^\beta}$ among all probability vectors $b$.
Define
\begin{IEEEeqnarray}{rCl}
\set{J} &\triangleq& \{i \in \{1,\ldots,m\} : b_i^* > 0\},\IEEEeqnarraynumspace\\
\Gamma &\triangleq& \frac{1 - \sum_{i \in \set{J}} p_i}{1 - \sum_{i \in \set{J}} \frac{1}{o_i}},\IEEEeqnarraynumspace\label{eq:partmoneydefgamma}
\end{IEEEeqnarray}
and for $i \in \{1,\ldots,m\}$,
\begin{IEEEeqnarray}{l}
\gamma_i \triangleq \max \vthinspace \Bigl\{0,\,\Gamma_{\vphantom{i}}^\frac{1}{\beta - 1} p_i^\frac{1}{1 - \beta} o_i^\frac{\beta}{1 - \beta} - \tfrac{1}{o_i}\Bigr\}.\IEEEeqnarraynumspace
\end{IEEEeqnarray}
Then, the following claims hold:
\begin{enumerate}
\item \label{clm:partmoneyreggammapositive} The quantity $\Gamma$ is well-defined and satisfies $\Gamma > 0$.
\item \label{clm:partmoneyregbi} For all $i \in \{1,\ldots,m\}$,
\begin{IEEEeqnarray}{l}
b_i^* = \gamma_i \vthinspace b_0^*.\IEEEeqnarraynumspace
\end{IEEEeqnarray}
\item \label{clm:partmoneyregbzero} The quantity $b_0^*$ satisfies
\begin{IEEEeqnarray}{l}
b_0^* = \frac{1}{1 + \sum_{i = 1}^m \gamma_i}.\IEEEeqnarraynumspace\label{eq:partmoneybzerostar}
\end{IEEEeqnarray}
In particular, $b_0^* > 0$.
\end{enumerate}
\end{Theorem}

\clmref{clm:partmoneyregbi} implies that for all $i \in \{1,\ldots,m\}$, $b_i^* > 0$ if and only if $p_i \vthinspace o_i > \Gamma$.
Assuming without loss of generality that $p_1 \vthinspace o_1 \ge p_2 \vthinspace o_2 \ge \ldots \ge p_m \vthinspace o_m$, the set $\set{J}$ thus has a special structure:
it is either empty or equal to $\{1,2,\ldots,k\}$ for some integer $k$.
To maximize $\frac{1}{\beta} \log \E{S_0^\beta}$, the following procedure can be used:
for every $\set{J}$ with the above structure, compute the corresponding $b$ according to \eqref{eq:partmoneydefgamma}--\eqref{eq:partmoneybzerostar};
and from these $b$'s, take one that maximizes $\frac{1}{\beta} \log \E{S_0^\beta}$.
This procedure leads to an optimal solution: an optimal solution $b^*$ exists because we are optimizing a continuous function over a compact set, and $b^*$ corresponds to a set $\set{J}$ that will be considered by the procedure.

\section{Proofs}
\label{sec:proofs}

\begin{proof}[Proof~of~\thmref{thm:allmoneyregular}]
We first show the maximization claim.
The only term on the RHS of \eqref{eq:thmallregsplit} that depends on $b$ is $-D_{1 - \beta}(g\|b)$.
Because $1 - \beta > 0$, this term is maximized if and only if $b = g$ \cite[Theorem~8]{VanErvenHarremoes}.

We now show \eqref{eq:thmallregsplit}.
By the definition of $S$,
\begin{IEEEeqnarray}{l}
\frac{1}{\beta} \log \E{S^\beta} = \frac{1}{\beta} \log \sum_{i = 1}^m p_i^{\vphantom{\beta}} \vthinspace o_i^\beta \vthinspace b_i^\beta.\IEEEeqnarraynumspace\label{eq:thmallregproofa}
\end{IEEEeqnarray}
For every $i \in \{1,\ldots,m\}$,
\begin{IEEEeqnarray}{rCl}
p_i^{\vphantom{\beta}} \vthinspace o_i^\beta \vthinspace b_i^\beta &=& \mleft[\vthinspace p_i^\frac{1}{1 - \beta} \vthinspace o_i^\frac{\beta}{1 - \beta}\mright]^{1 - \beta} \cdot b_i^\beta\IEEEeqnarraynumspace\label{eq:thmallregproofd}\\
&=& \mleft[\vthinspace\sum_{j = 1}^m p_j^{\frac{1}{1 - \beta}} o_j^{\frac{\beta}{1 - \beta}}\mright]^{1 - \beta} \cdot g_i^{1 - \beta} \vthinspace b_i^\beta,\IEEEeqnarraynumspace\label{eq:thmallregproofe}
\end{IEEEeqnarray}
where \eqref{eq:thmallregproofe} follows from \eqref{eq:thmallregdefgi}.
From \eqref{eq:thmallregproofa} and \eqref{eq:thmallregproofe} we obtain
\begin{IEEEeqnarray}{rCl}
\IEEEeqnarraymulticol{3}{l}{\frac{1}{\beta} \log \E{S^\beta}}\IEEEeqnarraynumspace\nonumber\\[-1ex]\qquad
&=& \frac{1 - \beta}{\beta} \log \sum_{j = 1}^m p_j^{\frac{1}{1 - \beta}} o_j^{\frac{\beta}{1 - \beta}} + \frac{1}{\beta} \log\sum_{i = 1}^m g_i^{1 - \beta} \vthinspace b_i^\beta\IEEEeqnarraynumspace\label{eq:thmallregproofh}\\
&=& \frac{1 - \beta}{\beta} \log \sum_{j = 1}^m p_j^{\frac{1}{1 - \beta}} o_j^{\frac{\beta}{1 - \beta}} - D_{1 - \beta}(g\|b)\IEEEeqnarraynumspace\label{eq:thmallregproofi}\\
&=& \log c + \frac{1 - \beta}{\beta} \log \sum_{j = 1}^m p_j^{\frac{1}{1 - \beta}} r_j^{\frac{-\beta}{1 - \beta}} - D_{1 - \beta}(g\|b)\IEEEeqnarraynumspace\label{eq:thmallregproofj}\\[0.5ex]
&=& \log c + D_{\!\frac{1}{1 - \beta}}(p\|r) - D_{1 - \beta}(g\|b),\IEEEeqnarraynumspace\label{eq:thmallregproofk}
\end{IEEEeqnarray}
where \eqref{eq:thmallregproofi} follows from identifying the R\'enyi divergence ($g$ and $b$ are probability vectors);
\eqref{eq:thmallregproofj} follows from \eqref{eq:defc} and \eqref{eq:defr}; and
\eqref{eq:thmallregproofk} follows from identifying the R\'enyi divergence ($p$ and $r$ are probability vectors).
This proves \eqref{eq:thmallregsplit}.
\end{proof}

\begin{proof}[Proof~of~\propref{prop:allmoneyzero}]
Equation \eqref{eq:propallzeroa} holds because
\begin{IEEEeqnarray}{rCl}
\lim_{\beta \to 0} \frac{1}{\beta} \log \E{S^\beta} &=& \lim_{\beta \to 0} \log \mleft[\vthinspace\sum_{i = 1}^m p_i^{\vphantom{\beta}} \vthinspace (o_i \vthinspace b_i)^\beta\mright]^\frac{1}{\beta}\IEEEeqnarraynumspace\label{eq:propallzeroproofa}\\
&=& \log \prod_{i = 1}^m (o_i \vthinspace b_i)^{p_i}\IEEEeqnarraynumspace\label{eq:propallzeroproofb}\\
&=& \sum_{i = 1}^m p_i \log \vthinspace (o_i \vthinspace b_i)\IEEEeqnarraynumspace\label{eq:propallzeroproofc}\\
&=& \E{\log S},\IEEEeqnarraynumspace\label{eq:propallzeroproofd}
\end{IEEEeqnarray}
where \eqref{eq:propallzeroproofa} follows from the definition of $S$, and
\eqref{eq:propallzeroproofb} holds because in the limit as $\beta$ tends to zero, the power mean tends to the geometric mean since $p$ is a probability vector \cite[Problem~8.1]{CSMC}.
Equation \eqref{eq:propallzerob} is proved in \cite[Section~10.3]{Moser}.
\end{proof}

\begin{proof}[Proof~of~\propref{prop:allmoneydegen}]
Inequality \eqref{eq:propalldegen} holds because
\begin{IEEEeqnarray}{rCl}
\frac{1}{\beta} \log \E{S^\beta} &=& \frac{1}{\beta} \log \sum_{i = 1}^m p_i^{\vphantom{\beta}} \vthinspace o_i^\beta \vthinspace b_i^\beta\IEEEeqnarraynumspace\label{eq:propalldegenproofa}\\
&\le& \frac{1}{\beta} \log \sum_{i = 1}^m p_i^{\vphantom{\beta}} \vthinspace o_i^\beta \vthinspace b_i^{\vphantom{\beta}}\IEEEeqnarraynumspace\label{eq:propalldegenproofb}\\
&\le& \frac{1}{\beta} \log \sum_{i = 1}^m b_i \cdot \max_{j \in \{1,\ldots,m\}} \bigl(\vthinspace p_j^{\vphantom{\beta}} \vthinspace o_j^\beta\bigr)\IEEEeqnarraynumspace\label{eq:propalldegenproofc}\\
&=& \frac{1}{\beta} \log \max_{j \in \{1,\ldots,m\}} \bigl(\vthinspace p_j^{\vphantom{\beta}} \vthinspace o_j^\beta\bigr)\IEEEeqnarraynumspace\label{eq:propalldegenproofd}\\
&=& \log \max_{j \in \{1,\ldots,m\}} \bigl(\vthinspace p_j^{1 / \beta} o_j^{\vphantom{1 / \beta}}\bigr),\IEEEeqnarraynumspace\label{eq:propalldegenproofe}
\end{IEEEeqnarray}
where \eqref{eq:propalldegenproofa} follows from the definition of $S$;
\eqref{eq:propalldegenproofb} holds because $b_i \in [0,1]$ and $\beta \ge 1$; and
\eqref{eq:propalldegenproofd} holds because $b$ is a probability vector.
It is easy to see that \eqref{eq:propalldegen} holds with equality if $b$ is chosen according to \eqref{eq:propalldegenoptb}.
\end{proof}

\begin{proof}[Proof~of~\propref{prop:allmoneyposinf}]
Equation \eqref{eq:propallposinfa} holds because
\begin{IEEEeqnarray}{rCl}
\lim_{\beta \to +\infty} \frac{1}{\beta} \log \E{S^\beta} &=& \lim_{\beta \to +\infty} \log \mleft[\vthinspace\sum_{i = 1}^m p_i^{\vphantom{\beta}} \vthinspace (b_i \vthinspace o_i)^\beta\mright]^\frac{1}{\beta}\IEEEeqnarraynumspace\label{eq:propallmoneyposinfa}\\*[0.5ex]
&=& \log \max_{i \in \{1,\ldots,m\}} b_i \vthinspace o_i,\IEEEeqnarraynumspace\label{eq:propallmoneyposinfb}
\end{IEEEeqnarray}
where \eqref{eq:propallmoneyposinfa} follows from the definition of $S$, and
\eqref{eq:propallmoneyposinfb} holds because in the limit as $\beta$ tends to $+\infty$, the power mean tends to the maximum since $p$ is a probability vector \cite[Chapter~8]{CSMC}.
Inequality \eqref{eq:propallposinfb} holds because $b_i \le 1$ for $i \in \{1,\ldots,m\}$.
It is easy to see that \eqref{eq:propallposinfb} holds with equality if $b$ is chosen according to \eqref{eq:propallposinfoptb}.
\end{proof}

\begin{proof}[Proof~of~\propref{prop:allmoneyneginf}]
Equation \eqref{eq:propallneginfa} holds because
\begin{IEEEeqnarray}{rCl}
\lim_{\beta \to -\infty} \frac{1}{\beta} \log \E{S^\beta} &=& \lim_{\beta \to -\infty} \log \mleft[\vthinspace\sum_{i = 1}^m p_i^{\vphantom{\beta}} \vthinspace (b_i \vthinspace o_i)^\beta\mright]^\frac{1}{\beta}\IEEEeqnarraynumspace\label{eq:propallmoneyneginfa}\\*[0.5ex]
&=& \log \min_{i \in \{1,\ldots,m\}} b_i \vthinspace o_i,\IEEEeqnarraynumspace\label{eq:propallmoneyneginfb}
\end{IEEEeqnarray}
where \eqref{eq:propallmoneyneginfa} follows from the definition of $S$, and
\eqref{eq:propallmoneyneginfb} holds because in the limit as $\beta$ tends to $-\infty$, the power mean tends to the minimum since $p$ is a probability vector \cite[Chapter~8]{CSMC}.

We show \eqref{eq:propallneginfb} by contradiction.
Assume that there exists a probability vector $b$ such that $\min_{i \in \{1,\ldots,m\}} b_i \vthinspace o_i > c$, i.e.,
\begin{IEEEeqnarray}{l}
b_i \vthinspace o_i > c\IEEEeqnarraynumspace\label{eq:propallmoneyneginfe}
\end{IEEEeqnarray}
for all $i \in \{1,\ldots,m\}$.
Then,
\begin{IEEEeqnarray}{rCl}
1 &=& \sum_{i = 1}^m b_i\IEEEeqnarraynumspace\label{eq:propallmoneyneginfh}\\
&>& \sum_{i = 1}^m \frac{c}{o_i}\IEEEeqnarraynumspace\label{eq:propallmoneyneginfi}\\
&=& 1,\IEEEeqnarraynumspace\label{eq:propallmoneyneginfj}
\end{IEEEeqnarray}
where \eqref{eq:propallmoneyneginfh} holds because $b$ is a probability vector;
\eqref{eq:propallmoneyneginfi} follows from \eqref{eq:propallmoneyneginfe}; and
\eqref{eq:propallmoneyneginfj} follows from the definition of $c$.
Because $1 > 1$ is impossible, such a $b$ cannot exist, which proves \eqref{eq:propallneginfb}.

It is easy to see that \eqref{eq:propallneginfb} holds with equality if $b_i = c / o_i$ for all $i \in \{1,\ldots,m\}$.
Conversely, if \eqref{eq:propallneginfb} holds with equality, then for all $i \in \{1,\ldots,m\}$,
\begin{IEEEeqnarray}{l}
b_i \vthinspace o_i \ge c.\IEEEeqnarraynumspace\label{eq:propallmoneyneginfl}
\end{IEEEeqnarray}
We claim that \eqref{eq:propallmoneyneginfl} holds with equality for all $i \in \{1,\ldots,m\}$.
Indeed, if this were not the case, then there would exist a $j \in \{1,\ldots,m\}$ for which $b_j \vthinspace o_j > c$, so \eqref{eq:propallmoneyneginfh}--\eqref{eq:propallmoneyneginfj} would hold, which would lead to a contradiction.
Hence, if \eqref{eq:propallneginfb} holds with equality, then $b_i = c / o_i$ for all $i \in \{1,\ldots,m\}$.
\end{proof}

\begin{proof}[Proof~of~\thmref{thm:sideinfo}]
We first show the maximization claim.
The only term on the RHS of \eqref{eq:thmsidesplit} that depends on $b_{X|Y}$ is $-D_{1 - \beta}(g_{X|Y} \vthinspace g_Y\|b_{X|Y} \vthinspace g_Y)$.
Because $1 - \beta > 0$, this term is maximized if and only if $b_{X|Y} \vthinspace g_Y = g_{X|Y} \vthinspace g_Y$ \cite[Theorem~8]{VanErvenHarremoes}.
By our assumptions that $p(y) > 0$ for all $y \in \set{Y}$ and $o(x) > 0$ for all $x \in \set{X}$, we have $g(y) > 0$ for all $y \in \set{Y}$.
Consequently, $b_{X|Y} \vthinspace g_Y = g_{X|Y} \vthinspace g_Y$ if and only if $b_{X|Y} = g_{X|Y}$.

We now show \eqref{eq:thmsidesplit}.
By the definition of $\tilde{S}$,
\begin{IEEEeqnarray}{l}
\frac{1}{\beta} \log \E{\tilde{S}^\beta} = \frac{1}{\beta} \log \sum_{x,y} p(x,y) \vthinspace o(x)^\beta \vthinspace b(x|y)^\beta.\IEEEeqnarraynumspace\label{eq:sideinfoa}
\end{IEEEeqnarray}
From \eqref{eq:thmsidedefgxy} and \eqref{eq:thmsidedefy} we obtain that for every $(x,y) \in \set{X} \times \set{Y}$,
\begin{IEEEeqnarray}{rCl}
\IEEEeqnarraymulticol{3}{l}{p(x,y) \vthinspace o(x)^\beta \vthinspace b(x|y)^\beta}\IEEEeqnarraynumspace\nonumber\\\qquad
&=& \sum_{y'} p(y') \mleft[\sum_{x'} p(x'|y')^\frac{1}{1 - \beta} o(x')^\frac{\beta}{1 - \beta}\mright]^{1 - \beta}\hphantom{.}\IEEEeqnarraynumspace\nonumber\\
&& \hfill \binaryemptyleft \cdot g(y) \vthinspace g(x|y)^{1 - \beta} \vthinspace b(x|y)^\beta.\IEEEeqnarraynumspace\label{eq:sideinfod}
\end{IEEEeqnarray}
Now, \eqref{eq:thmsidesplit} holds because
\begin{IEEEeqnarray}{rCl}
\IEEEeqnarraymulticol{3}{l}{\frac{1}{\beta} \log \E{\tilde{S}^\beta}}\IEEEeqnarraynumspace\nonumber\\[-1ex]\quad
&=& \frac{1}{\beta} \log \sum_{y'} p(y') \mleft[\sum_{x'} p(x'|y')^\frac{1}{1 - \beta} o(x')^\frac{\beta}{1 - \beta}\mright]^{1 - \beta}\IEEEeqnarraynumspace\nonumber\\
&& \binaryemptyleft + \frac{1}{\beta} \log \sum_{x,y} \bigl[g(x|y) \vthinspace g(y)\bigr]^{1 - \beta} \bigl[b(x|y) \vthinspace g(y)\bigr]^\beta\IEEEeqnarraynumspace\label{eq:sideinfog}\\
&=& \log c + \frac{1}{\beta} \log \sum_{y'} p(y') \mleft[\sum_{x'} p(x'|y')^\frac{1}{1 - \beta} r(x')^\frac{-\beta}{1 - \beta}\mright]^{1 - \beta}\IEEEeqnarraynumspace\hspace{-1.1em}\nonumber\\
&& \binaryemptyleft + \frac{1}{\beta} \log \sum_{x,y} \bigl[g(x|y) \vthinspace g(y)\bigr]^{1 - \beta} \bigl[b(x|y) \vthinspace g(y)\bigr]^\beta\IEEEeqnarraynumspace\label{eq:sideinfoh}\\
&=& \log c + D_{\!\frac{1}{1 - \beta}}(p_{X|Y}\|r_X|p_Y)\IEEEeqnarraynumspace\nonumber\\
&& \binaryemptyleft - D_{1 - \beta}(g_{X|Y} \vthinspace g_Y\|b_{X|Y} \vthinspace g_Y),\IEEEeqnarraynumspace\label{eq:sideinfoi}
\end{IEEEeqnarray}
where \eqref{eq:sideinfog} follows from plugging \eqref{eq:sideinfod} into \eqref{eq:sideinfoa} and using the fact that $g(y) = g(y)^{1 - \beta} \vthinspace g(y)^\beta$;
\eqref{eq:sideinfoh} follows from \eqref{eq:defrx}; and
\eqref{eq:sideinfoi} follows from identifying the conditional R\'enyi divergence and the (unconditional) R\'enyi divergence.
\end{proof}

\begin{proof}[Proof~of~\propref{prop:conddivprop}]
We first show \eqref{eq:conddivnonzero}.
If $\alpha \in (0,1)$, then H\"older's inequality implies that for all $y \in \set{Y}$,
\begin{IEEEeqnarray}{l}
\sum_x p(x|y)^\alpha \vthinspace q(x|y)^{1 - \alpha} \le \! \mleft[\sum_x p(x|y)\mright]^\alpha \mleft[\sum_x q(x|y)\mright]^{1 - \alpha}\!.\IEEEeqnarraynumspace\label{eq:conddivpropa}
\end{IEEEeqnarray}
The RHS of \eqref{eq:conddivpropa} equals one, so
\begin{IEEEeqnarray}{C}
\log \sum_y p(y) \mleft[\sum_x p(x|y)^\alpha \vthinspace q(x|y)^{1 - \alpha}\mright]^\frac{1}{\alpha} \le 0,\IEEEeqnarraynumspace\label{eq:conddivprope}
\end{IEEEeqnarray}
which implies \eqref{eq:conddivnonzero} because $\frac{\alpha}{\alpha - 1} < 0$.
If $\alpha > 1$, then the inequalities in \eqref{eq:conddivpropa} and \eqref{eq:conddivprope} are reversed; since now $\frac{\alpha}{\alpha - 1} > 0$, \eqref{eq:conddivnonzero} holds also in this case.

We now show \eqref{eq:conddivlefullcond}.
If $\alpha > 1$, then \eqref{eq:conddivlefullcond} holds because
\begin{IEEEeqnarray}{rCl}
\IEEEeqnarraymulticol{3}{l}{\frac{\alpha}{\alpha - 1} \log \sum_y p(y) \mleft[\sum_x p(x|y)^\alpha \vthinspace q(x|y)^{1 - \alpha}\mright]^\frac{1}{\alpha}}\IEEEeqnarraynumspace\nonumber\\\quad
&\le& \frac{\alpha}{\alpha - 1} \log \mleft[\sum_y p(y) \sum_x p(x|y)^\alpha \vthinspace q(x|y)^{1 - \alpha}\mright]^\frac{1}{\alpha}\IEEEeqnarraynumspace\label{eq:conddivproph}\\
&=& \frac{1}{\alpha - 1} \log \sum_{x,y} [p(x|y) \vthinspace p(y)]^\alpha [q(x|y) \vthinspace p(y)]^{1 - \alpha},\IEEEeqnarraynumspace\label{eq:conddivpropi}
\end{IEEEeqnarray}
where \eqref{eq:conddivproph} follows from Jensen's inequality because $z \mapsto z^\frac{1}{\alpha}$ is a concave function on $\reals_{\ge 0}$,
and \eqref{eq:conddivpropi} holds because $p(y) = p(y)^\alpha \vthinspace p(y)^{1 - \alpha}$.
If $\alpha \in (0,1)$, then $z \mapsto z^\frac{1}{\alpha}$ is convex, so Jensen's inequality is reversed; because $\frac{\alpha}{\alpha - 1} < 0$, \eqref{eq:conddivproph} and thus \eqref{eq:conddivlefullcond} hold also in this case.
\end{proof}

\begin{proof}[Proof~of~\propref{prop:uncondleconddiv}]
If $\alpha > 1$, then \eqref{eq:uncondleconddiva} holds because
\begin{IEEEeqnarray}{rCl}
\IEEEeqnarraymulticol{3}{l}{D_\alpha(p_X\|r_X)}\IEEEeqnarraynumspace\nonumber\\\quad
&=& \frac{\alpha}{\alpha - 1} \log \mleft\{\sum_x \mleft[p(x) \vthinspace r(x)^\frac{1 - \alpha}{\alpha}\mright]^\alpha\mright\}^\frac{1}{\alpha}\IEEEeqnarraynumspace\\
&=& \frac{\alpha}{\alpha - 1} \log \mleft\{\sum_x \mleft[\sum_y p(y) \vthinspace p(x|y) \vthinspace r(x)^\frac{1 - \alpha}{\alpha}\mright]^\alpha\mright\}^\frac{1}{\alpha}\IEEEeqnarraynumspace\\
&\le& \frac{\alpha}{\alpha - 1} \log \sum_y p(y) \mleft\{\sum_x \mleft[p(x|y) \vthinspace r(x)^\frac{1 - \alpha}{\alpha}\mright]^\alpha\mright\}^\frac{1}{\alpha}\IEEEeqnarraynumspace\label{eq:uncondleconddivb}\\
&=& D_\alpha(p_{X|Y}\|r_X|p_Y),\IEEEeqnarraynumspace
\end{IEEEeqnarray}
where \eqref{eq:uncondleconddivb} follows from the Minkowski inequality \cite[III 2.4 Theorem 9]{MeansAndInequalities}.
If $\alpha \in (0,1)$, then the Minkowski inequality is reversed; since now $\frac{\alpha}{\alpha - 1} < 0$, \eqref{eq:uncondleconddivb} and thus \eqref{eq:uncondleconddiva} hold also in this case.
\end{proof}

\begin{proof}[Proof~of~\propref{prop:fairsuperfair}]
Set $b_0' = 0$ and $b_i' = r_i \vthinspace b_0 + b_i$ for all $i \in \{1,\ldots,m\}$.
Then, $\sum_{i = 0}^m b_i' = 1$, and for $i \in \{1,\ldots,m\}$,
\begin{IEEEeqnarray}{rCl}
o_i \vthinspace b_i' &=& c \vthinspace b_0 + o_i \vthinspace b_i\IEEEeqnarraynumspace\\
&\ge& b_0 + o_i \vthinspace b_i,\IEEEeqnarraynumspace\label{eq:fairsuperfairb}
\end{IEEEeqnarray}
where \eqref{eq:fairsuperfairb} holds because $c \ge 1$.
It is not difficult to see that \eqref{eq:fairsuperfairb} implies \eqref{eq:fairsuperfairbprime}.
\end{proof}

\begin{proof}[Proof~of~\thmref{thm:partmoneyregular}]
In the \hyperlink{prf:partmoneyregular}{Appendix}.
\end{proof}

\newpage
\appendix

\begin{proof}[Proof~of~\thmref{thm:partmoneyregular}]
\hypertarget{prf:partmoneyregular}{}
The proof is based on the Karush--Kuhn--Tucker conditions.

Assume first that $\beta \in (0,1)$, and define the function $\tau$ from the set of probability vectors to the set of real numbers as
\begin{IEEEeqnarray}{l}
\tau(b) \triangleq \sum_{i = 1}^m p_i \vthinspace (b_0 + b_i \vthinspace o_i)^\beta.\IEEEeqnarraynumspace\label{eq:partmoneyregdeftau}
\end{IEEEeqnarray}
Since $\smash{\frac{1}{\beta}} > 0$ and since the logarithm is an increasing function, maximizing $\frac{1}{\beta} \log \E{S_0^\beta}$ is equivalent to maximizing $\tau$.

Observe that $\tau$ is concave, so by the Karush--Kuhn--Tucker conditions \cite[Theorem~4.4.1]{Gallager}, it is maximized by a probability vector $b$ if and only if there exists a $\lambda \in \reals$ such that (i) for all $i \in \{0,\ldots,m\}$ with $b_i > 0$,
\begin{IEEEeqnarray}{l}
\frac{\partial \tau}{\partial b_i} (b) = \lambda,\IEEEeqnarraynumspace
\end{IEEEeqnarray}
and (ii) for all $i \in \{0,\ldots,m\}$ with $b_i = 0$,
\begin{IEEEeqnarray}{l}
\frac{\partial \tau}{\partial b_i} (b) \le \lambda.\IEEEeqnarraynumspace\label{eq:partmoneyregkktpreb}
\end{IEEEeqnarray}
From now on, we use the following notation: (i) and (ii) hold simultaneously if and only if for all $i \in \{0,\ldots,m\}$,
\begin{IEEEeqnarray}{l}
\frac{\partial \tau}{\partial b_i} (b) \begin{cases} = \lambda & \text{if $b_i > 0$,}\\
\le \lambda & \text{if $b_i = 0$.}\end{cases}\IEEEeqnarraynumspace\label{eq:partmoneyregkkta}
\end{IEEEeqnarray}
Dividing both sides of \eqref{eq:partmoneyregkkta} by $\beta > 0$ and defining $\mu \triangleq \frac{\lambda}{\beta}$, we obtain
\begin{IEEEeqnarray}{l}
\frac{1}{\beta} \cdot \frac{\partial \tau}{\partial b_i} (b) \begin{cases} = \mu & \text{if $b_i > 0$,}\\
\le \mu & \text{if $b_i = 0$.}\end{cases}\IEEEeqnarraynumspace\label{eq:partmoneyregkktd}
\end{IEEEeqnarray}
Evaluating \eqref{eq:partmoneyregkktd} leads to
\begin{IEEEeqnarray}{l}
\sum_{i = 1}^m p_i \vthinspace (b_0 + b_i \vthinspace o_i)^{\beta - 1} \begin{cases} = \mu & \text{if $b_0 > 0$,}\\
\le \mu & \text{if $b_0 = 0$,}\end{cases}\IEEEeqnarraynumspace\label{eq:partmoneyregkktg}
\end{IEEEeqnarray}
and for all $i \in \{1,\ldots,m\}$,
\begin{IEEEeqnarray}{l}
p_i \vthinspace o_i \vthinspace (b_0 + b_i \vthinspace o_i)^{\beta - 1} \begin{cases} = \mu & \text{if $b_i > 0$,}\\
\le \mu & \text{if $b_i = 0$.}\end{cases}\IEEEeqnarraynumspace\label{eq:partmoneyregkkth}
\end{IEEEeqnarray}

Assume now that $\beta < 0$, and define $\tau$ as in \eqref{eq:partmoneyregdeftau}.
Then, maximizing $\frac{1}{\beta} \log \E{S_0^\beta}$ is equivalent to minimizing $\tau$ because now $\smash{\frac{1}{\beta}} < 0$.
The function $\tau$ is convex, so the inequality in \eqref{eq:partmoneyregkktpreb} is reversed.
Dividing by $\beta < 0$ again reverses the inequalities, so \eqref{eq:partmoneyregkktd}--\eqref{eq:partmoneyregkkth} hold also if $\beta < 0$.

Let $\beta \in (-\infty,0) \cup (0,1)$, and let $b^*$ be a probability vector that maximizes $\frac{1}{\beta} \log \E{S_0^\beta}$.
By the above discussion, \eqref{eq:partmoneyregkktg} and \eqref{eq:partmoneyregkkth} are satisfied by $b^*$ for some $\mu \in \reals$.
The LHS of \eqref{eq:partmoneyregkktg} is positive, so $\mu > 0$.
We now show that for all $i \in \{1,\ldots,m\}$,
\begin{IEEEeqnarray}{l}
b_i^* = \max \vthinspace \biggl\{0,\,\biggl[\frac{p_i \vthinspace o_i^\beta}{\mu}\biggr]^\frac{1}{1 - \beta} - \frac{b_0^*}{o_i}\biggr\}.\IEEEeqnarraynumspace\label{eq:partmoneyregbimaxa}
\end{IEEEeqnarray}
Fix $i \in \{1,\ldots,m\}$.
If $b_i^* > 0$, then \eqref{eq:partmoneyregkkth} implies
\begin{IEEEeqnarray}{l}
b_i^* = \biggl[\frac{p_i \vthinspace o_i^\beta}{\mu}\biggr]^\frac{1}{1 - \beta} - \frac{b_0^*}{o_i}.\IEEEeqnarraynumspace\label{eq:partmoneyregbimaxd}
\end{IEEEeqnarray}
Because $b_i^* > 0$, the RHS of \eqref{eq:partmoneyregbimaxd} is positive, which proves \eqref{eq:partmoneyregbimaxa} in the case $b_i^* > 0$.
If $b_i^* = 0$, then \eqref{eq:partmoneyregkkth} implies
\begin{IEEEeqnarray}{l}
\biggl[\frac{p_i \vthinspace o_i^\beta}{\mu}\biggr]^\frac{1}{1 - \beta} - \frac{b_0^*}{o_i} \le 0.\IEEEeqnarraynumspace\label{eq:partmoneyregbimaxg}
\end{IEEEeqnarray}
The LHS of \eqref{eq:partmoneyregbimaxg} is nonpositive, so \eqref{eq:partmoneyregbimaxa} holds also if $b_i^* = 0$.

We show next that $b_j^* = 0$ for some $j \in \{1,\ldots,m\}$.
For a contradiction, assume that $b_i^* > 0$ for all $i \in \{1,\ldots,m\}$.
Then,
\begin{IEEEeqnarray}{rCl}
\sum_{i = 1}^m p_i \vthinspace (b_0^* + b_i^* \vthinspace o_i)^{\beta - 1} &=& \mu \cdot \sum_{i = 1}^m \frac{1}{o_i}\IEEEeqnarraynumspace\label{eq:partmoneyregspeciald}\\
&>& \mu,\IEEEeqnarraynumspace\label{eq:partmoneyregspeciale}
\end{IEEEeqnarray}
where \eqref{eq:partmoneyregspeciald} follows from \eqref{eq:partmoneyregkkth}, and
\eqref{eq:partmoneyregspeciale} holds because $c < 1$ by assumption.
But this is impossible: \eqref{eq:partmoneyregspeciale} contradicts \eqref{eq:partmoneyregkktg}.

Now, let $j \in \{1,\ldots,m\}$ be such that $b_j^* = 0$.
Because $p_j$ and $o_j$ are positive, this implies $b_0^* > 0$ by \eqref{eq:partmoneyregbimaxa}.
Thus, \eqref{eq:partmoneyregkktg} implies
\begin{IEEEeqnarray}{l}
\sum_{i = 1}^m p_i \vthinspace (b_0^* + b_i^* \vthinspace o_i)^{\beta - 1} = \mu.\IEEEeqnarraynumspace\label{eq:partmoneyregmua}
\end{IEEEeqnarray}
Splitting the sum on the LHS of \eqref{eq:partmoneyregmua} depending on whether $b_i^* = 0$ or $b_i^* > 0$, we obtain
\begin{IEEEeqnarray}{rCl}
\mu &=& \sum_{i \in \set{J}} p_i \vthinspace \biggl\{o_i \cdot \biggl[\frac{p_i \vthinspace o_i^\beta}{\mu}\biggr]^\frac{1}{1 - \beta}\biggr\}^{\beta - 1} + \sum_{i \notin \set{J}} p_i \vthinspace (b_0^*)^{\beta - 1}\IEEEeqnarraynumspace\label{eq:partmoneyregmud}\\
&=& \mu \sum_{i \in \set{J}} \frac{1}{o_i} + (b_0^*)^{\beta - 1} \biggl[1 - \sum_{i \in \set{J}} p_i\biggr],\IEEEeqnarraynumspace\label{eq:partmoneyregmue}
\end{IEEEeqnarray}
where \eqref{eq:partmoneyregmud} follows from \eqref{eq:partmoneyregbimaxa}.
Rearranging \eqref{eq:partmoneyregmue}, we obtain
\begin{IEEEeqnarray}{l}
\mu \biggl[1 - \sum_{i \in \set{J}} \frac{1}{o_i}\biggr] = (b_0^*)^{\beta - 1} \biggl[1 - \sum_{i \in \set{J}} p_i\biggr].\IEEEeqnarraynumspace\label{eq:partmoneyregmuh}
\end{IEEEeqnarray}
Because $b_j^* = 0$, $j \notin \set{J}$, so $1 - \sum_{i \in \set{J}} p_i > 0$.
Additionally, $\mu > 0$ and $b_0^* > 0$, so $1 - \sum_{i \in \set{J}} \frac{1}{o_i} > 0$.
This proves \clmref{clm:partmoneyreggammapositive}, because it implies that $\Gamma$ is well-defined and positive.

By the definition of $\Gamma$, \eqref{eq:partmoneyregmuh} implies
\begin{IEEEeqnarray}{l}
\mu = \Gamma \vthinspace (b_0^*)^{\beta - 1}.\IEEEeqnarraynumspace\label{eq:partmoneyregmuk}
\end{IEEEeqnarray}
Plugging \eqref{eq:partmoneyregmuk} into \eqref{eq:partmoneyregbimaxa} establishes \clmref{clm:partmoneyregbi}, i.e., that for all $i \in \{1,\ldots,m\}$,
\begin{IEEEeqnarray}{l}
b_i^* = \gamma_i \vthinspace b_0^*.\IEEEeqnarraynumspace\label{eq:partmoneyregbigammai}
\end{IEEEeqnarray}

We move on to \clmref{clm:partmoneyregbzero}.
Because $b^*$ is a probability vector,
\begin{IEEEeqnarray}{rCl}
1 &=& b_0^* + \sum_{i = 1}^m b_i^*\IEEEeqnarraynumspace\label{eq:partmoneyregbzgammaa}\\
&=& b_0^* \vthinspace \biggl[1 + \sum_{i = 1}^m \gamma_i\biggr],\IEEEeqnarraynumspace\label{eq:partmoneyregbzgammab}
\end{IEEEeqnarray}
where \eqref{eq:partmoneyregbzgammab} follows from \eqref{eq:partmoneyregbigammai}.
Now, \eqref{eq:partmoneyregbzgammab} implies \clmref{clm:partmoneyregbzero}.
\end{proof}


\begin{thebibliography}{99}
\bibitem{Kelly} J.~L.~Kelly, ``A new interpretation of information rate,'' \emph{Bell Syst. Tech. J.}, vol.~35, no.~4, pp.~917--926, Jul.~1956.
\bibitem{CoverThomas} T.~M.~Cover and J.~A.~Thomas, \emph{Elements of Information Theory}. 2nd ed. Hoboken, NJ, USA: John Wiley \& Sons, 2006.
\bibitem{CSMC} J.~M.~Steele, \emph{The Cauchy--Schwarz Master Class}. Cambridge: Cambridge Univ. Press, 2004.
\bibitem{MeansAndInequalities} P.~S.~Bullen, \emph{Handbook of Means and Their Inequalities}. Dordrecht, The Netherlands: Kluwer Academic Publishers, 2003.
\bibitem{CampbellA} L.~L.~Campbell, ``A coding theorem and R\'enyi's entropy,'' \emph{Inf. Control}, vol.~8, no.~4, pp.~423--429, Aug.~1965.
\bibitem{CampbellB} L.~L.~Campbell, ``Definition of entropy by means of a coding problem,'' \emph{Z. Wahrscheinlichkeitstheorie verw. Geb.}, vol.~6, no.~2, pp.~113--118, Jun.~1966.
\bibitem{ExpMoments} N.~Merhav, ``On optimum strategies for minimizing the exponential moments of a loss function,'' in \emph{Proc. 2012 IEEE Int. Symp. Inf. Theory}, 2012, pp.~140--144.
\bibitem{Soklakov} A.~N.~Soklakov, ``Economics of disagreement -- financial intuition for the R\'enyi divergence,'' 2018, arXiv:1811.08308v4.
\bibitem{RenyiDivergence} A.~R\'enyi, ``On measures of entropy and information,'' in \emph{Proc. 4th Berkeley Symp. Math. Statist. and Probability}, vol.~1, 1961, pp.~547--561.
\bibitem{VanErvenHarremoes} T.~van~Erven and P.~Harremo\"es, ``R\'enyi divergence and Kullback--Leibler divergence,'' \emph{IEEE Trans. Inf. Theory}, vol.~60, no.~7, pp.~3797--3820, Jul.~2014.
\bibitem{VerduAlphaMutual} S.~Verd\'u, ``$\alpha$-mutual information,'' in \emph{Proc. 2015 Inf. Theory and Appl. Workshop (ITA)}, 2015, pp.~1--6.
\bibitem{Moser} S.~M.~Moser, ``Information Theory (Lecture Notes),'' version 6.2, 2018. [Online]. Available: \url{http://moser-isi.ethz.ch/scripts.html}
\bibitem{Gallager} R.~G.~Gallager, \emph{Information Theory and Reliable Communication}. Hoboken, NJ, USA: John Wiley \& Sons, 1968.
\end{thebibliography}
\end{document}